# A repository of automatic GUI test patterns in Android applications: Specification and Analysis using Alloy modeling language

Fatemeh Mosayeb[1] ،Shohreh Ajoudanian [2, 3*]
[1] Faculty of Computer Engineering, Islamic Azad University of Najafabad, Iran,
Fatememosayeb71@sco.iaun.ac.ir
[2] Faculty of Computer Engineering, Islamic Azad University of Najafabad, Iran,
shajoudanian@pco.iaun.ac.ir
[3] Big Data Research Center, Najafabad Branch, Islamic Azad University, Najafabad, Iran
*Corresponding author

**Abstract**

The software industry aims to provide customers with quality software. Testing software is a critical and sensitive stage in ensuring software quality. Due to the increasing popularity of mobile devices, the use of Android applications has increased. Almost all are equipped with Graphical User Interface (GUI) to interact with users or systems. GUI is the most common tool to communicate with modern software. Therefore, the perfect GUI is a GUI that ensures the safety, strength, and usability of the whole software system. The GUI testing is a vital stage in ensuring the product quality because the GUI is the user's first impression and the final view of the final product. This paper has proposed a new technique to promote the model-based test efficiency using Alloy modeling language. The findings showed that this approach needs less configuration and modeling time than previous methods. Moreover, using GUI patterns may decrease errors and violations.

## 1. Introduction

Advances in the field of mobile devices' processing power have led to their increased popularity and capabilities. The existing software distribution model for all operating systems is through online application stores or markets [1]. These online stores have allowed programmers to effortlessly and cheaply distribute applications among plenty of users. Based on the ever-increasing popularity of Android operating systems, a lack of accurate security has caused many concerns. Additionally, there are countless system and security failures reported in Android smartphones. Therefore, Android application developers predominantly use software tests as a necessary step to examine software maintenance and errors [2]. Studies show that the Android system has more than 80% market share in the world's cell phone industry [1]. Applications or software constitute the most substantial part of an Android phone. Statistics show that more than 3.8 million applications exist for the Android platform [2].

The most crucial part of Android software is its activity. Activities manage the user interface (UI) page and define all interactions between users and UI [3]. Considering the continuous progress of the Android system's software for practical use cases in the industry and economy and increased user expectations regarding promoted efficiency in terms of accuracy and speed, developers have been encouraged to create more superior and higher quality software. Among other tasks, the Android software test is one of the most fundamental procedures in developing high-quality software. Almost all Android applications are equipped with a Graphical User Interface (GUI) to interact with users or the system.

This study focuses on Android applications' GUI test. The GUI test is a vital step in ensuring product quality because the graphical interface is the starting point of the user's first impression and final view of the final product [4].

Generally, the variables investigated in this study include the number of detected failures and test patterns. This study attempts to answer the following questions:

**RQ1:** In contrast to other model-based GUI test approaches, what are the benefits of the proposed method?

**RQ2:** Are existing elements and connectors for GUI modeling error detectors sufficient and influential?

Our contributions are as follows:

- Developing a model to test applications and automatically producing test cases aiming to test GUI in Android applications in less time and cost,
- Accurate classification and description to eliminate existing defects in GUI of Android applications, and
- User interface test modeling in Android applications using Alloy modeling language.

This paper is organized as follows: in section 2, concepts related to software tests and the presented approach are defined. Section 3 discusses studies on software GUI. Section 4 is about the proposed method and introduces the structure of the suggested patterns. Section 5 analyzes and assesses the proposed approach. Section 6 shows information about the existing challenges. Finally, Section 7 concludes the paper.

## 2. Background

Every Android application consists of four components. Every component has a significant share in creating Android applications. These four components include activities, services, broadcast receivers, and content providers. 1) Activities provide UI and manage user interaction with the smartphone screen. 2) Services manage background processing of the application. 3) Broadcast receivers manage the relationship between the Android system and applications. 4) Content providers manage subjects related to data and database management [5].

Android application test is a challenging task with unresolved and specific problems. For instance, most developers are not familiar with the Android platform and consequently leave their programs vulnerable to new types of bugs.

Providing suitable methods for software tests facilitates bug detection. According to definitions, most GUI failures result from the difference between right and wrong behavioral descriptions of a GUI [6]. GUI failures may appear in various ways. In other words, they may include different types of GUI objects that can emerge in varied ways. These failures are classified based on domain and mode features; 1) the domain shows the type of involved GUI in the graphical failure, and 2) the mode shows how to display the error at the end of the GUI [7].

In particular, the domain feature equals one of the GUI objects used to run the application intended for the mobile platform. For instance, various GUI objects used to run GUI on the Android platform include Bottom, ContextMenu, Dialog, and TextView.

### 2.1. Alloy modeling language

Alloy language was introduced by MIT University's Software Design Department led by Daniel Jackson in 1997. The first version was limited to object modeling. This model gradually became a complete structural modeling language. Furthermore, it enjoys an accurate mathematical semantic based on relational algebra. Alloy displays its operands based on Transition System modes. However, in algebraic semantic relations, it is not entirely limited to displaying modes. Generally, Alloy is a popular modeling relational language that resembles programming languages class-wise. Moreover, in terms of semantics, it is based on first-order logic and z logic structure [8].

Alloy is used for model analysis and detecting inconsistencies in the design and recently has attracted a lot of attention in the research community. Alloy Analyzer can check the consistency of an object model expressed in Alloy, produce snapshots, perform operations, and check their properties. Alloy can have optional relationships and structural mechanisms to reuse model components [9].

Unlike a programming language, an Alloy model is declarative. Therefore it can explain the effect of behavior without providing its mechanism. This feature makes the production and analysis of very brief and detailed models possible. Also, Alloy can be analyzed automatically and be accepted as a control for the model.

In Alloy, constraints, and statements are defined as constraints on model elements. Alloy language is automatically supported by a constraints solver called Alloy analyzer. Users specify domains. In case an assertion violation is found, then it is not valid. However, if no case is found, the assertion may be invalid on a larger scale. One of the crucial characteristics of Alloy is that its behavior is based on numbers and relations. Phrases in Alloy are similar to the mathematical standard of relational logic. The necessary Alloy commands are as follows [8].

- Signatures: define a set in this language.

- Facts: state facts about modeling. Indeed, they focus on the constraints and facts that can be used in modeling.
- Fields: it describes a relationship between two sets.
- Predicates: templates for parametric constraints. They can be implemented elsewhere using parameter exploration or true or false evaluation.
- Functions: parametric formulas that are used elsewhere. Functions can be used elsewhere by using parameter exploration from parameter examples and evaluating a value.
- Assertions: constraints that want to follow the facts of a model. It is a relationship that its accuracy should be investigated, assuming the facts are within the model.

## 3. Related Work

This section investigates various approaches proposed in the context of the Android GUI test. A recording and playback tool called Reran for the Android platform is presented in [10]. The recording module records GUI and sensor event flow in this tool, and the playback tool accurately plays back these events. A versatile and lightweight recording and playback tool called VALERA was proposed for Android [11]. They used the sensor playback technique to increase the accuracy of finding existing bugs in applications under test. A platform and playback tool called Mosaic is presented for Android [12]. According to one approach, the application model under test can be used to generate a sequence of events for each purpose, which can be lines or branches in the source code. This method, first, manually presents a summary of the application event and then uses the generated model to generate a sequence of events [13]. They introduced Stoat as an autonomous method based on the model to promote the GUI test. It uses behavioral models of the software to repeat the test sequence towards the top cover and the sequence of various events [14].

ORBIT is another model extraction technique that creates a GUI model to test applications. ORBIT uses static analysis to detect user interface events in certain activities like mode changes [15]. An official model and architecture have been presented to test running applications. The system behavior under test in this model is specified as a labeled transmission system [16].

GUITAR is a GUI test framework to run JAVA and Microsoft Windows applications [17]. Negiun et al. described the architecture of GUITAR as a model-based framework and architecture to test the Android software [18]. In a model-based method, GUI was introduced by using Petri net technique called HPrTNs. HPrTNs are a higher class of Petri net [19].

An approach was introduced based on Finite State Machine (FSM) called GUITam to automatize all event flow graphs-based method tests [20]. This approach is concentrated on mobile phone applications. This study considers both GUI events and contextual events supported by mobile applications [21]. A model of applications is created using the requirements, a standard method in model-based tests. To this end, Interactive Cooperative Object (ICO) was identified as one of the modeling techniques that allows one to describe GUI complex behavior [22]. In one approach, a model-based test has been done to test a mobile application in the real world. This model shows the application from the user perspective in which nodes present perceived modes and edges show measures by the user [23].

In a method for automatic production, graphical events were presented in the Android applications test. In this article, a random algorithm was introduced for dynamicity in generating a sequence of events for the GUI test of the Android software [24]. iMPAct tool tests frequent behaviors or UI patterns in mobile applications. To test the software, they used test methods based on behavioral and graphical patterns repetition [25]. In similar applications, an approach has been suggested to extract user behavior patterns. As a result, costs to figure out new applications decrease [26]. They proposed a framework to test consistency between Android application components. Generally, according to their studies, Android applications communicate with other components through intents [27].

## 4. The Proposed Method

This section presents the proposed approach to help Android applications automatic graphical interface tests using GUI. Therefore, this section exposes a modeling and testing method that takes advantage of application interface patterns used to design GUI. In Fig. 1, the steps taken in this proposed approach are presented. The First steps are the collection phase, meta-model design, and collection of Android graphical interface patterns. Then, the next phase is devoted to the meta-model and Android graphical interface test patterns modeling in the Alloy modeling language.

Moreover, pattern-based tests in Alloy and the generation of Android graphical interface test cases are done. Finally, Alloy instances are analyzed to check that the constraints in the model are met.

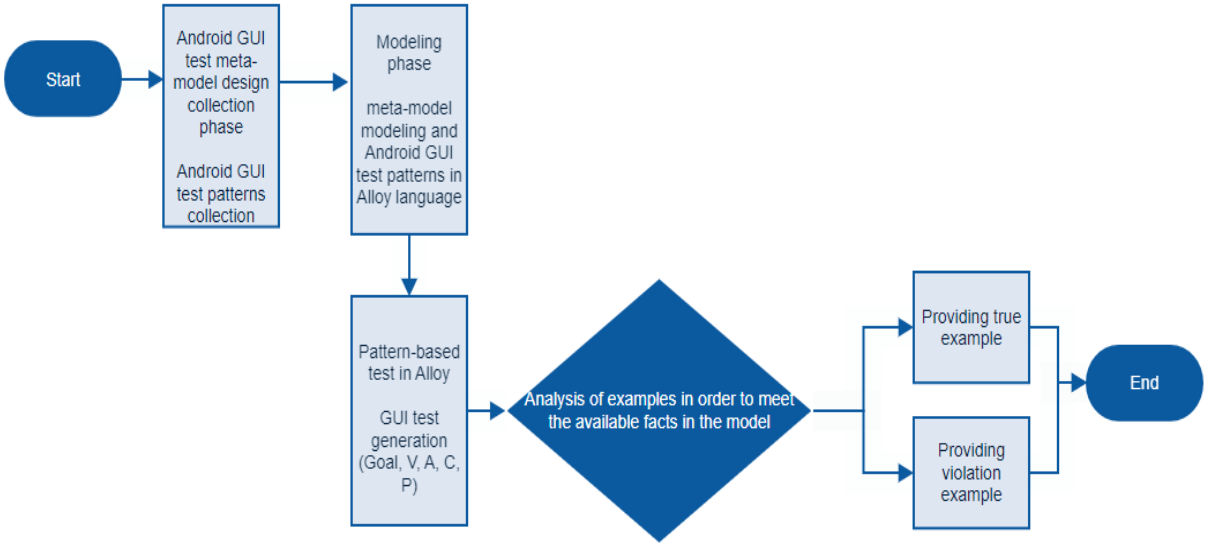

Figure 1. The proposed approach flowchart

**4.1. The meta-model of the proposed approach**

The pattern-based graphical interface test approach needs a GUI model to create a domain. This language reuses the existing elements or redevelops them to promote reusability. The proposed meta-model in this study is a DSL built to test a pattern-based graphical interface.

A model records all abstractions of the domain and specifies their relationships. Abstractions refer to the concepts in the context of the PBGT pattern-based test. In the proposed meta-model, these abstractions are the elements of UI test patterns, structural elements, and language connectors. A DSL model written in UML (Fig. 2) can be defined as a meta-model with a UML profile. As in Fig. 2, this meta-model inspired by [28] consists of elements and connections. The pink sections in Fig. 2 are related to the work done in this paper.

**Elements**: A component is an abstract entity that shows the available concepts in the field of PBGT. The model this meta-model starts with an Init and ends with an End. Both Init and End are abstract elements. As models grow sizable to deal with their growing complexity, structural techniques such as different hierarchical levels can be used. These techniques allow the use of Model A inside Model B, which shows the details of A inside B. That is, it is possible to have a nested structure such as what occurs in C and Java languages with structures such as modules. The form element structure has been created specifically for this purpose. A form (a structural element) includes a model (minor model) with Init and End elements.

Groups are also structural elements but do not have End and Init elements. Moreover, all elements inside the group are executed in the desired order. Furthermore, behavioral elements represent UI test patterns that define strategies to test UI patterns.

**Connectors**: A connector is a popular, expressive, and potent graphic symbol to indicate tasks. Markedly, these connectors define a set of time operators to combine tasks. The proposed meta-model defines three connectors: Sequences, Sequence WithMovedData, and Sequence WithDataPassing.

- **The Sequence connector** shows that the target element cannot be started before the source element expires.
- **The Sequence WithDataPassing connector** manifests a similar behavior to "Sequence" and shows that the target element received data from the source element.

- **The Sequence WithMovedData connector** has definitions similar to Sequence WithDataPassing. However, "attachment" is another type of relationship among elements. This relation shows that the target element depends on the source element set properties. For example, when the result is computational.
- The proposed meta-model includes additional facts that cannot be directly expressed in the UML language model. Along with DSL development, some primary views on language limitations progress over time. It has been decided to build the language model in Alloy and follow the repetitive structural trend to help determine the limits and constraints imposed on language. This language model is explained in the following.

### 4.2. Meta-model modeling with Alloy

Alloy is a lightweight formal language that supports structural and behavioral modeling. It is suitable for primary stages of software development that allow proper abstraction detection of software.

Alloy is based on relational logic that combines first-order logic measurement criteria with relational calculus operators. Moreover, Alloy allows the expression of complex structural and behavioral facts. The advantages of Alloy are simplicity and accuracy.

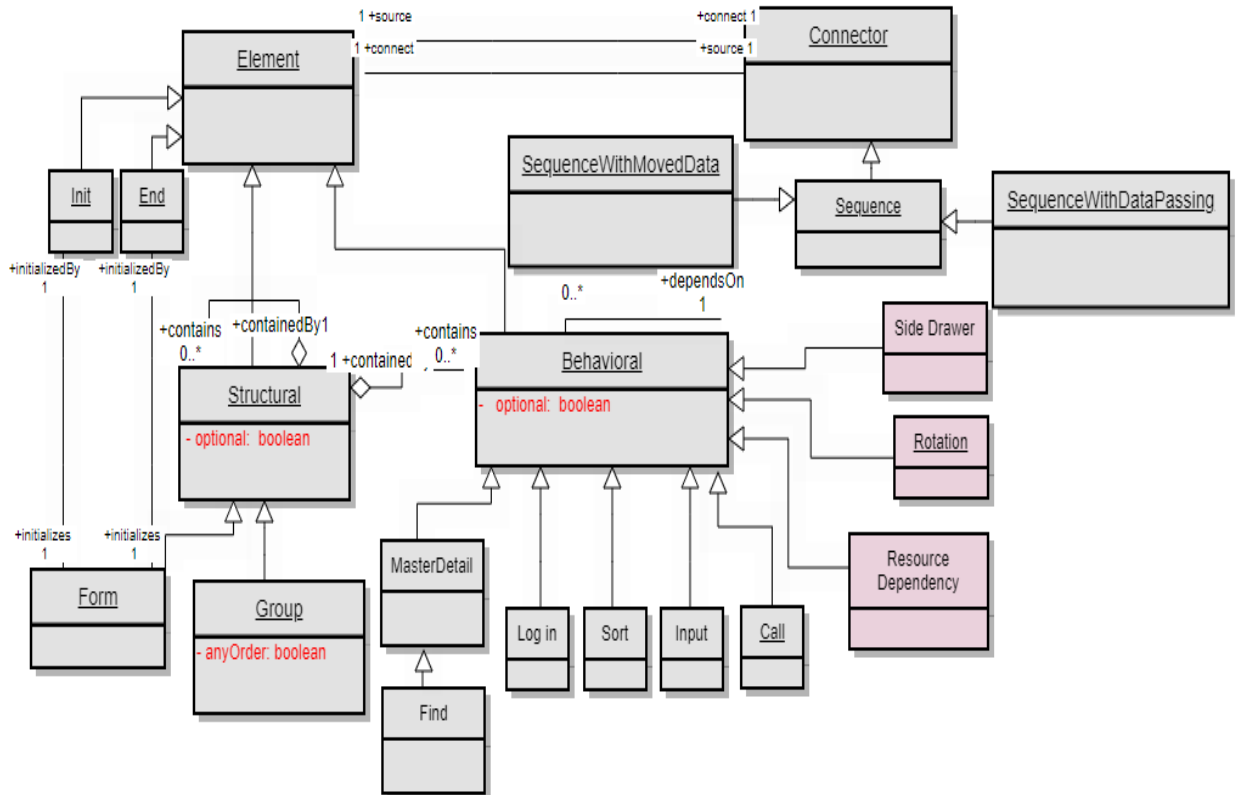

Figure 2. The proposed approach meta-model inspired by [28]

The meta-model in Fig. 2 is modeled using Alloy to produce instances that should validate GUI test models. In this paper, Alloy 4 is used for specification. It has a library of predefined modules. In the first line of Listing 1, the ordering module is defined for action order. Markedly, in the third line, a set of elements are defined. In lines 3 to 7, the Init, End, Behavior, and Structural elements are defined. In line 8, it is declared that each structural element is related to several innerStructs (related to structural element connections at various abstract levels). In line 9, each behavioral element can be related to one or more innerBehavior (related to connecting structural elements with behavioral elements). A subset of structural elements is defined in lines 10 and 11 with at least two behavioral patterns. Also, line 12 is a subset of structural elements with a Start and an End element. Line 15 is a subset of the form.

Line 16 is the parent function that its inputs are elements. This function has structural components of innerStruct and innerBehavior. In line 18, constraints and facts are pointed out. In lines 19 and 20, all elements in the model have a parent, but the model itself lacks it. In line 21, it is stated that for all Ss that are structural elements in the model, a relationship of s exists in the model and innerStructs.

```
 1: open util/ordering[Action]
 2: open util/boolean
 3: abstract sig Element{}
 4: sig Init extends Element{}
 5: sig End extends Element{}
 6: abstract sig Behaviour extends Element{}
 7: abstract sig Structural extends Element{
 8: innerStructs:set Structural,
 9:  innerBehaviour:some Behaviour
}
10: sig Group extends Structural{
11: {#innerBehaviour>1}}
12: sig Form extends Structural{
13: init: one Init,
14:  end: one End
}
15: sig Model extends Form{}
16: fun Parent[e:Element]:Structural{
17: innerStructs.e+innerBehaviour.e+init.e+end.e
}
18: fact{
19: all   e:Element-Model   | one Parent[e]
20: no Parent[Model]
21: all s:Structural-Model | s in Model.^ innerStructs}
```

Listing 1. Android applications GUI test meta-model modeling in Alloy

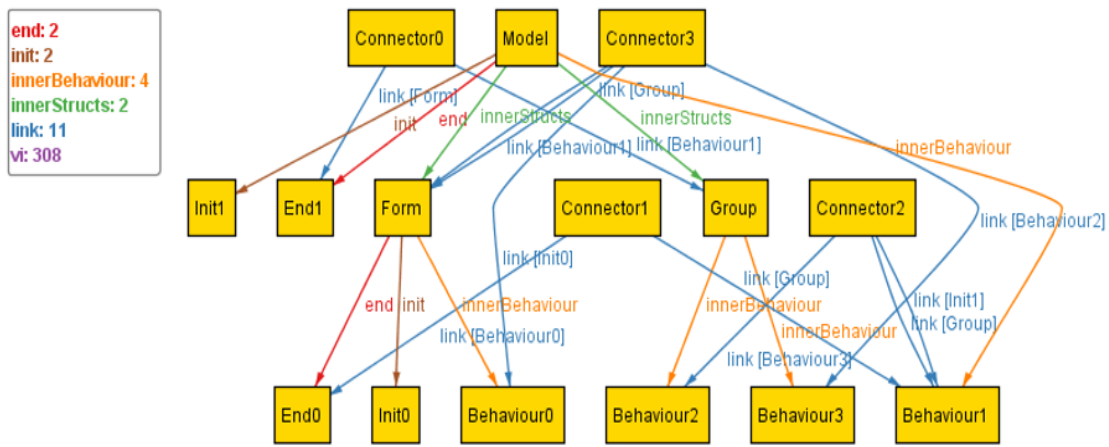

Figure 3. An example of GUI test meta-model modeling in Alloy

Fig. 3 shows an example of GUI test modeling with two hierarchical levels (Model and Form). This model has an Init (Init1), an End (End1), a Behavior of UI display pattern (Behavior), a Form, and a Group. Also, the Form has an Init (Init0), an End (End0), and a UI behavior test pattern. This group has two test patterns, UI Behavior2 and Behavior2. This example allows us to have correct model specifications (with designed facts).

Then, we developed the model for connector descriptions and language facts (Listing 2) to ensure each connector only connects two structural elements (Model, Form, and Group).

Facts and constraints about the model are defined in line 1. In lines 2 and 3, it is stated that for all forms, connections from the End element and innerStructs and innerBehavior exist, and the Start element will be Connector.link (related to the connection of connectors to structural or behavioral elements). In lines 4 and 5, for all member Fs, a connection form exists from the Start element, innerBehavior, and innerStructs until the End element of the same Form. A set of connectors are defined in line 6. In line 7, it is declared that each element can only be connected to one element. In line 8, the Link.element can never be connected to Link[Element], meaning connectors cannot connect an element of themselves. In line 9, no connection exists between the Link.element and the End element. In line 10, no connection exists between Link[Element] and the Start element. In line 11, it is stated that no connection exists between Link[Element] (related to the connection between the connector and structural or behavioral elements), Link.element (related to the connection between connector and element but with a different combination), and the model element. In line 12, at last, a connection exists between Link[Element], Link.element, and InitEnd. In line 13, it is stated that no connection exists between Link[Element] and elements that their parents belong to a structural element.

```
1:fact {
2: all f:Form | f.end + f.innerBehaviour + f.innerStructs
3: in f.init.^(Connector.link)
4: all f:Form | f.init + f.innerBehaviour + f.innerStructs
5: in f.end.^(~(Connector.link))}
6: sig Connector{
7: link:Element -> lone Element
}{
8: no link.Element & link[Element]
9:  no link.Element&End
10: no link[Element] & Init
11: no (link[Element]+link.Element)& Model
12: lone(link[Element]+link.Element)&(Init+End)
13: no e:link[Element] | Parent[link.e] !=Parent[e]
}
```

Listing 2. Alloy model with proposed meta-model connectors

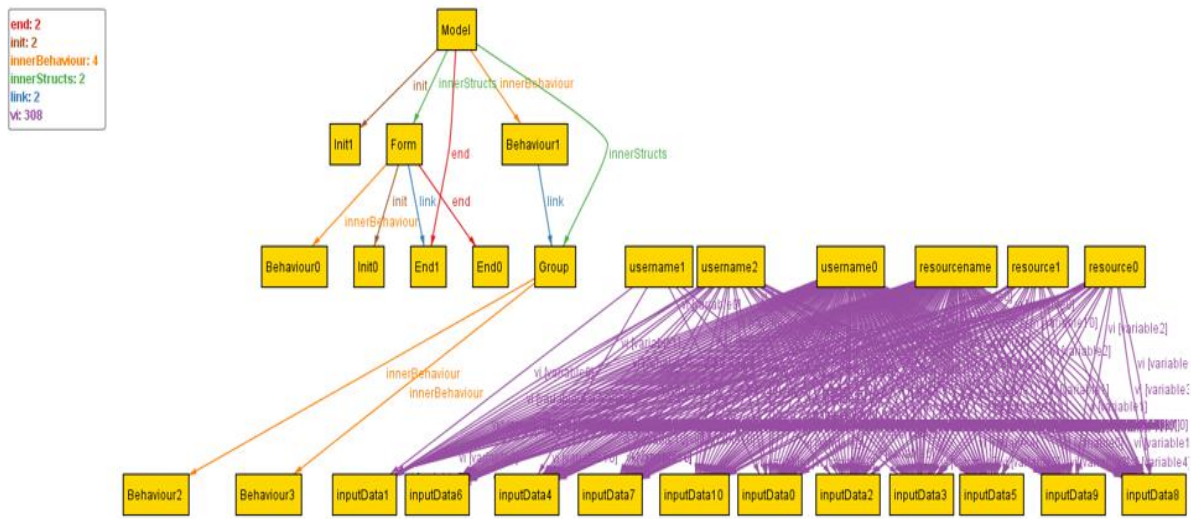

Figure 4. An example of alloy model about connectors

The language facts described for the proposed meta-model are as follows:

**LC1:** A connector cannot connect more than one element to itself.

**LC2:** A connector cannot have Init as a source or destination.

**LC3:** An Init element cannot directly connect to an End element.

**LC4:** Two elements cannot be connected more than once by connectors simultaneously.

**LC5:** Two elements can be connected to each other only if they belong to the same structural element (Model, Form, and Group).

**LC6:** For all elements within a form, at least one path exists from Init to End that passes that element.

### 4.3. Android graphical interface pattern modeling in Alloy

In this approach, the patterns combine the items listed in Table 2. Accordingly, Goal identifies the pattern, and the Precondition defines test execution conditions. *A* is a sequence of measures to execute the test, and *C* shows the test status. *V* contains customizable preset information for the pattern (username and password match in the login/password template).

Table 2. Pattern structure definition

| Pattern= <G, V, A, C, P> | |
|---|---|
| **Goal** | Test goal |
| **Variable** | Test input variables |
| **Action** | A sequence of ongoing actions |
| **Check** | A set of checks performed during the test |
| **Precondition** | Defining the prerequisites of the conditions under which the test can be performed |

These patterns are defined as follows:

$$G[configuration] : P \rightarrow A[V] \rightarrow C$$

To configure any goal (*G*), if the precondition (*P*) is confirmed, a sequence of actions (*A*) with input values (*V*) is executed. In the end, a set of checks (*C*) is performed.

In the following subsections, the proposed patterns for GUI test are presented.

#### 4.3.1. Side drawer pattern

A side drawer is a panel that enters from left to right or vice versa and displays contents. The user moves his hand from left to right and clicks on the action bar button to see this menu.

This pattern is defined according to Table 3. In line 1, a side drawer pattern with its inputs is defined. Here, the goal is the availability of the side drawer pattern. A series of actions are performed, such as reading the page or pressing a specific button. Finally, it is checked whether the pattern exists or not.

Table 3. Side drawer pattern definition

| Pattern structure = <G, V, A, C, P> | | Alloy code |
|---|---|---|
| **Goal** | Side Drawer exists | 1: **pred** SideDrawerPattern [g: Goal, v: Vari, c: Checks, a: Action]{ |
| **Variable** | {} | 2:  **g** in SIDEDRAWER |
| **Action** | [read screen] | 3:  **v** in Vari |
| **Check** | {"side drawer exists and is hidden"} | 4:  **c** in full |
| **Precondition** | {true} | 5:  **a** in readscreen or a in opensidedrawer or a in readscreen  } |

### 4.3.2. Rotation pattern

Android devices have two possible screen orientations: portrait and landscape. When the device rotates, the display orients, and its design updates. However, according to the Android guidelines, testing involves two main aspects for the developers to consider: UI customized code to manage these changes, and no user input information should be lost.

Therefore, testing this pattern is used to check display rotation behavior and whether display elements and inputs are displayed before rotation or not.

The rotation pattern is defined according to Table 4. In line 1, test preconditions with inputs are defined. In this pattern, the goal is to orient the display. While rotation, it is checked whether the pattern is executed correctly or not.

Table 4. Rotation pattern definition

| Pattern structure = <G, V, A, C, P> | | Alloy code |
|---|---|---|
| Goal | ROTATIONISPOSSIBLE | 1: **pred** OrientationPattern [g: Goal, v: Vari, c: Checks, a: Action]{ <br> 2:  **g** in ROTATIONISPOSIBLE <br> 3:  **v** in Vari <br> 4:  **c** in mainC <br> 5:  **a** in readscreen or a in rotatescreen or a in readscreen or a in scrollscreen <br> } |
| Variable | {} | |
| Action | [read screen] | |
| Check | itispossibletorotatescreen | |
| Precondition | {true} | |

### 4.3.3. Resources dependency pattern

Many Android applications use external sources like GPS or WiFi. Moreover, many are dependent on the accessibility of these sources. Accordingly, checking that the applications do not fail when the source is unavailable is necessary. Available sources stop using this pattern, and no failure occurs if the application is confirmed. The resource dependency pattern is defined according to Table 5. Test execution preconditions and inputs are defined in line 1. The goal of this test is the scenario when the sources are unavailable, and the application does not execute correctly. The input variables of the test are sources. Actions include shutting down these sources and reading the display. Also, the pattern performance is checked.

Table 5. Resources dependency pattern definition

| Pattern structure = <G, V, A, C, P> | | Alloy code |
|---|---|---|
| Goal | Resource in use | 1: **pred** ResourceDependencyPattern [g: Goal, v: Vari, c: Checks, a: Action]{ <br> 2:  **g** in notCrash <br> 3:  **v** in resource **v** in resourcename <br> 4:  **c** in applicationC <br> 5:  **a** in readscreen or a in turnresourceoff or a in readscreen <br> } |
| Variable | {"resource", resource_name} | |
| Action | [read resource status] | |
| Check | {"resource is being used by the app"} | |
| Precondition | {true} | |

### 4.3.4. Tab-Scroll pattern

A Tab-Scroll allows switchability among different views in applications.

Tab-Scroll is defined according to Table 6. In line 1, this pattern is defined with its execution preconditions. The goal is to test the presence of this pattern. Moreover, different application parts are checked, and measures are performed, such as switching between different application tabs.

Table 6. Tab-Scroll pattern definition

| Pattern structure = <G, V, A, C, P> | | Alloy code |
|---|---|---|
| Goal | PRESENCE | 1: **pred** TabScrollPattern [g: Goal, v: Vari, c: Checks, a: Action]{ |
| Variable | {} | 2:   g in PRESENCE |
| Action | [read screen] | 3:   v in Vari |
| Check | TabsP | 4:   c in tabsP |
| Precondition | {true} | 5:   a in readS<br>} |

### 4.3.5. Input UI Test pattern

This pattern should be used to test the input behavior about the validity or invalidity of input data. Input UI Test pattern is defined according to Table 7. In line 1, the preconditions of this pattern with inputs are specified. The goal is to test valid and invalid data. The test variables are inputs. While executing the pattern, a series of inputs should be presented. Finally, true or false inputs are checked by an inbox or a label.

Table 7. Input UI Test pattern definition

| Pattern structure = <G, V, A, C, P> | | Alloy code |
|---|---|---|
| Goal | Valid data and Invalid data | 1: **pred** InputUITestPattern[g: Goal, v: Vari, c: Checks, a: Action]{ |
| Variable | {input} | 2:   g in INPVD or g in INPID |
| Action | [provide input] | 3:   v in input |
| Check | {message box", "label","error provider"} | 4:   c in errorP  or c in messageB or c in lable |
| | | 5:   a in a in provideI |
| Precondition | {true} | } |

### 4.3.6. Login UI Test pattern

This pattern should be used to test identification. The goal is to authenticate a valid username and password.

Login UI Test is defined according to Table 8. In line 1, a pattern with its execution preconditions is defined. Its inputs are username and password, and the actions include inserting username and password and pressing the button. Finally, whether the display is changed or an error is displayed is checked.

Table 8. Login UI Test pattern definition

| Pattern structure = <G, V, A, C, P> | | Alloy code |
|---|---|---|
| Goal | Valid login and Invalid login | 1: **pred** LoginUITestPattern[g: Goal, v: Vari, c: Checks, a: Action]{ |
| Variable | {username, password} | |
| Action | [provide *username*; provide *password*;press *submit*] | 2:   g in LGVAL or g in LGINV |
| | | 3:   v in username or v in password |
| Check | hange to page X", "pop-up error Y", "same page", "label K"} | 4:   c in changeP or c in popupE or c in sameP |
| | | 5:   a in provideU or a in provideP or a in pressS |
| Precondition | {true} | } |

### 4.3.7. Master/Detail UI Test

This pattern should be used to test UI patterns that include two parts: major and minor. It should be applied so that with major changes, details change too.

Master/Detail UI Test is defined according to Table 9. In line 1, the pattern precondition and its inputs are defined. The goal of this pattern is major/minor changes.

Table 9. Master/Detail UI Test

| Pattern structure = <G, V, A, C, P> | | Alloy code |
|---|---|---|
| Goal | Change master | 1: **pred** MasterDetailUITestPattern[g: Goal, v: Vari, c: Checks, a: Action]{ |
| Variable | {master,detail} | |
| Action | [select master] | 2:   **g** in MD |
| Check | { "detail has value X", "detail, does not have value X", "detail is empty} | 3:   **v** in master or **v** in detail |
| | | 4:   **c** detaiX or **c** in detailnotX or **c** in detailS  5:   **a** in selectMaster |
| Precondition | {true} | } |

### 4.3.8. Find UI Test

This pattern should be used when a person wants to test the search results following an action. The goal is the confirmation that states the search results are consistent with expectations (acceptable values are found).

Find UI Test pattern is defined according to Table 10. In line 1, pattern precondition and inputs are defined. The goal is to find values. Inputs are given while test execution and actions aim to provide these inputs. Finally, whether the results are correct or not is checked, and an estimation is performed.

Table 10. Find UI Test pattern

| Pattern structure = <G, V, A, C, P> | | Alloy code |
|---|---|---|
| Goal | "Value found" and "Value not found" | 1: **pred** FindUITestPattern[g: Goal, v: Vari, c: Checks, a: Action]{ |
| Variable | V: {v1,...,vN} where N is defined during configuration time by the user/tester; | 2:   **g**  in FNDVF or **g** in FNDNF |
| | | 3:   **v**  in vari |
| Action | [provide v1,... , provide vN] | 4:   **c**  emptyS or **c** in Xelements or **c** in notelementX or **c** in resultXisY |
| Check | {"result is an empty set", "result has X elements", "results do not have element X", "result has element X in line Y", "results have more than X elements", "results have less than X elements"} | 5:   **a**  in provide |
| | | } |
| Precondition | {true} | |

### 3.4.9. Sort UI Test

This pattern is used to check whether the result is customized according to the selected sorting criteria or not. The idea behind this pattern is that user interfaces contain detachable items/elements such as tables and lists.

Sort UI Test pattern is defined according to Table 11. In line 1, pattern precondition and inputs are defined. The goal is sorting based on ascending or descending modes. Inputs are given to the user while pattern execution. Actions aim to provide these inputs. In the end, it is checked whether the sorting result is acceptable and if the element X in position Y has the value of z or not.

Table 11. Sort UI Test pattern definition

| Pattern structure = <G, V, A, C, P> | Alloy code |
|---|---|

| Goal | Ascending and descending |
|---|---|
| Variable | {(v1, c1),...,(vN, cN)} |
| Action | [provide (v1, c1),... , provide (vN, c1)]; |
| Check | {"element from field X in position Y has value Z", "elements (v) with a given criteria (c) are sorted ac Y"}. |
| Precondition | {true} |

```
1: pred SortUITestPattern [g:
   Goal, v: Vari, c: Checks, a:
   Action]{
2:   g in SRTASC or g in SRTDESC
3:   v in Vari
4:   c in elementProv
5:   a in provide
   }
```

### 4.3.10. Call UI Test

This UI test pattern is used to check the performance of related writings. When configuring, the user only should check and define its precondition. The Call UI Test pattern is defined according to Table 12. In line 1, preconditions of this pattern and inputs are defined. The goal is to check the correctness or incorrectness of the action. By pressing the button, it is checked that the same page is displayed or it has changed to another one.

Table 12. Call UI Test pattern definition

| Pattern structure = <G, V, A, C, P> | | Alloy code |
|---|---|---|
| Goal | Action invoked | |
| Variable | no input data is involved | |
| Action | [press] | |
| Check | {"pop-up message", "stay in the same page", "change to page X"}. | |
| Precondition | {true} | |

```
1: pred CallUITestPattern [g:
   Goal, v: Vari, c: Checks, a:
   Action]{
2:   g in CLAS or g in CLAF
3:   v in Vari
4:   c in popupM or c in stayP or c in changeP
5:   a in press
   }
```

## 5. Case Study

In this section, we validated the proposed approach using instantiating. A set of common Android applications are tested to evaluate the proposed approach.

### 5.1. Validation of the proposed approach using instantiating

The proposed meta-model is used to convert an XML file to Alloy. In instantiating, an Android UI is considered as an XML file input (Fig.15). The structure of the applied application in Android based on the XML file is hierarchical (Fig.16). This hierarchical structure consists of *View* and *ViewGroup*. Views are observable elements on the Android user interface, are visible to the end user, and the user interacts with them. In the proposed meta-model, these views are those structural elements such as buttons, text boxes, radio boxes, etc.

*ViewGroup* acts as storage in which several views exist and specify the layout or arrangement of the UI page. It can be *LinearLayout* or *ConstraintLayout,* and in the proposed meta-model, it is the *Group*. In the instantiating, an XML file has been used in which a vertical *LinearLayout* exists (Fig 17). Test patterns can be checked in this example.

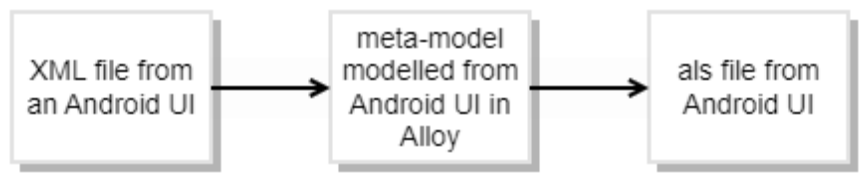

Figure 15. Android UI conversion to Alloy

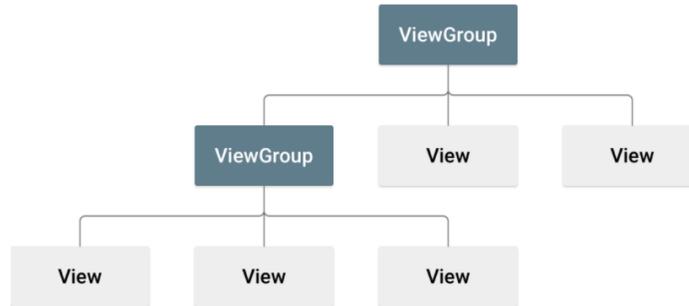

Figure 16. Android applications structure

```
<?xml version="1.0" encoding="utf-8"?>
<LinearLayout xmlns:android="http://schemas.android.com/apk/res/android"
        android:layout_width="match_parent"
        android:layout_height="match_parent"
        android:orientation="vertical" >
  <TextView android:id="@+id/text"
        android:layout_width="wrap_content"
        android:layout_height="wrap_content"
        android:text="Hello, I am a TextView" />
  <Button android:id="@+id/button"
        android:layout_width="wrap_content"
        android:layout_height="wrap_content"
        android:text="Hello, I am a Button" />
</LinearLayout
```

Figure 17. Android xml file

## 5.2. Analysis and Results

A set of common Android applications are gathered in this section to evaluate the proposed approach. The goal is to assess the pattern-based graphical interface test approach from the modeling perspective until the generated test items exit. The first stage to assess the PBGT approach is based on checking the configuration and test time. Finally, the configuration and modeling of the proposed approach are compared with other modeling approaches such as spec# and VAN4GUIM.

### 5.2.1 Time analysis

First, the time (minute) required to create and configure models using the three mentioned approaches is estimated. Table 13 shows the measured time for modeling and configuration.

Table 13. Comparing time required for configuration and test of the proposed approach with other two approaches

| Apps | Methods | | |
|---|---|---|---|
| | Spec# | VAN4GUIM | Proposed approach |
| App1 | M=420 C=56 | M=118 C=40 | M=60 C=3 |
| App2 | M=390 C=42 | M=160 C=30 | M=40 C=1.5 |
| App3 | M=210 C=35 | M=117 C=29 | M=45 C=1.5 |
| App4 | M=390 C=53 | M=222 C=34 | M=50 C=3.5 |
| App5 | M=455 | M=310 | M=108 |
|      | C=66   | C=48   | C=2   |
| App6 | M=211  | M=178  | M=55  |
|      | C=50   | C=42   | C=3   |
| App7 | M=301  | M=186  | M=40  |
|      | C=58   | C=33   | C=2   |

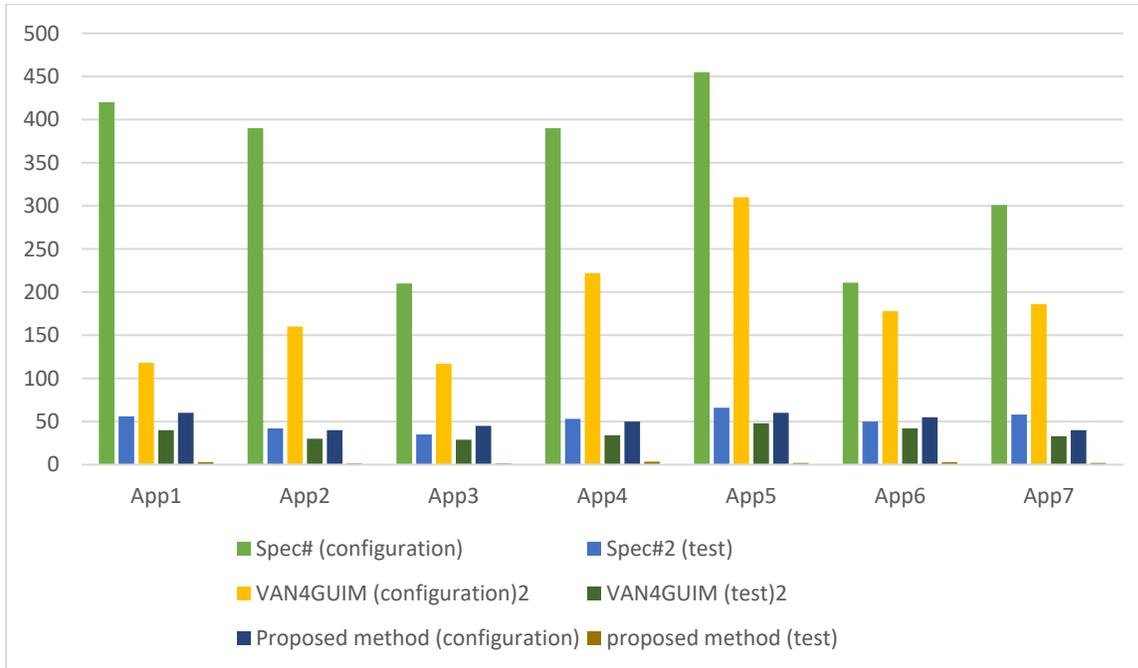

Figure 18. Comparing time required for configuration between the proposed approach and Spec# and VAN4GUIM

According to Table 13, the time required for the proposed approach's configuration and modeling is less than the other two approaches. For example, the modeling time for the App5 is 108 minutes. While for the same app, the required time of the other two approaches is more.

**5.3. Investigating dependent and independent variables**

Generally, this study's variables are classified into two groups: the number of test patterns or independent variables and the number of detected failures or dependent variables.

First, the number of patterns in each application should be specified to investigate independent and dependent variables. For example, the application providing banking services, among the ten graphical interface patterns defined in this study, has all the patterns except for rotation. For the graphical interface test, this application is generated without using patterns. However, by adding patterns to the model, failure cases decrease. Finally, according to Table 14, it is observed that with more patterns in the model, failures decrease and even reach zero.

Table 14. The relationship between number of patterns and failures in applications

| Application | Before using the patterns | using 2 patterns | Maximum use of pattern |
|-------------|---------------------------|------------------|------------------------|
| App1        | 5                         | 4                | 2                      |
| App2        | 7                         | 6                | 4                      |
| App3        | 4                         | 3                | 1                      |
| App4        | 5                         | 3                | 2                      |
| App5        | 8                         | 7                | 5                      |
| App6        | 7                         | 6                | 4                      |
| App7        | 7                         | 5                | 3                      |

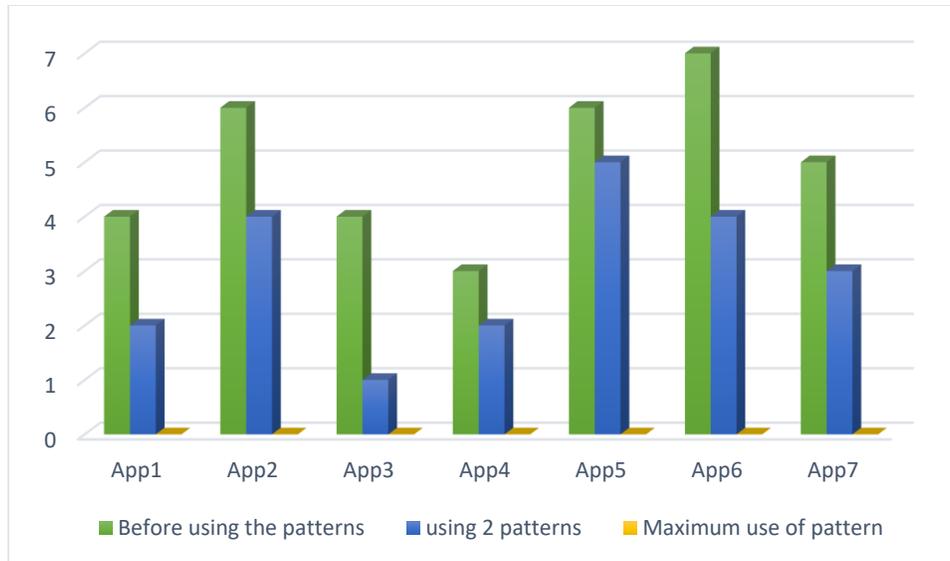

Figure 19. The relationship between number of patterns and violations

## 6. Limitations and Discussion

Over the last two decades, the GUI test has faced many challenges despite the availability of emergent technologies and tools. A summary of prominent findings and challenges impacting the GUI test and future studies are presented in this section:

- The main goal of the GUI test is action. However, reliability, execution, and security are minor goals.
- The most common method for the GUI test is the model-based method. However, different techniques are used in evaluating secondary results.

In what follows, top challenges in this area and future directions are presented:

- **Test processes and approaches:** The GUI test for mobile applications is challenging. This difficulty is due to the dynamicity of the mobile applications world that constantly includes new development concepts such as activities and interactions. For this purpose, several new techniques such as model-based, search-based, and random methods stand. Although the results of these studies have been encouraging, several questions remain unanswered. Questions like what are the best systematic approaches and processes of GUI test and in which applications they are used.
- **Test case generation:** Today, the complexity of GUI tests needs process automation. Many studies are devoted to this topic. However, several questions remain, such as how to generate the best and most effective test cases.
- **Understanding the application behavior:** Investigations show that features or various events detection in GUI test helps to access high coverage. Exploration and understanding of the behavior of these applications are demanding. Diverse techniques are proposed for the effective exploration of events in GUI. However, the increasing complexity of mobile applications, understanding application behavior, and exploring event sequences for effective testing remains challenging for researchers.
- **Minimizing and prioritizing test:** Several researchers suggested techniques to minimize test set with positive performance. But this open question needs to be addressed. Moreover, GUI test in many situations is a demanding and time-consuming process. Therefore, it is vital to know how to prioritize GUI test cases. It is crucial to discover various techniques in this area and employ them to prioritize tests.
- **Create/maintain model-based tests:** The time and cost of creating and maintaining the model with a GUI application are volatile. It is similar to maintaining test scripts to prevent application changes from failing test execution. An option to avoid time-consuming tasks is manually creating models and automatically passing the SUT. While this process allows the rapid production of thousands of test cases, it may not be effective in finding failures or providing SUT coverage. This problem is valid for the GUI test because test execution at the GUI level requires much more time than at the unit level.

Three model-based GUI test approaches of Spec#, VAN4GUIM, and the proposed approach are selected to address RQ1. The time required to create and configure models was measured in three approaches. Table 14 shows time measurement based on the minute for modeling (as M) and configuration (as C). Table 14 results indicate that the time required for modeling and configuring the proposed approach is less than the other two. For RQ3, a failure example is generated without using the patterns to test GUI. However, by adding the patterns to the model, these cases decreased. Finally, according to Table 15, it is observed that the more patterns in the model, the fewer violation examples exist.

## 7. Conclusion

The purpose of this article was to model UI tests in Android applications with Alloy modeling language. The main contribution of this study was using Alloy language features and presenting the proposed approach. These features also include saving time in modeling and configuration. Although many approaches were proposed for UI tests and modeling, the available approaches needed more time and had less accuracy because of their lack of attention to different modeling methods.

Furthermore, the available approaches comprehensively view the modeling methods and UI test. They have also selected the modeling method with Alloy modeling language and the first-order relational logic approach. So far, no Alloy model has been proposed to improve UI tests for Android applications, which constitutes the main advantage of this study over others. The PBGT approach is a model-based GUI test method that aims to improve GUI test strategies. Markedly, this approach introduces UI test patterns that have defined general test strategies and can execute different UI patterns. The UI test patterns show behavioral elements in the meta-model. In general, creating a meta-model for PBGT allows the tester to describe models that describe test goals. Therefore, PBGT includes a set of UI test patterns. PBGT is supported by a modeling environment and a uniform test consisting of several components that empower the approach and facilitate the use.

The PBGT-based approach fills the gap between design and quality guarantee. UI patterns used to design GUI for applications provide a systematic approach to testing several UI pattern implementations to ensure their performance. This method can improve the existing GUI models and tests by decreasing the actions needed to build models and directing the test focus towards test goals.

Compared to other model-based test approaches, PBGT requires less time for modeling and configuration. The most important course of action for the future are as follows:

- More UI test patterns can be added to the study and the proposed approach to keep going and develop GUI test approaches.
- A vast set of UI patterns can be detected in various operating systems by adding more rules and facilities and improving test efficiency and quality. About the PBGT approach, this approach cannot test iOS applications. Therefore, one of the future works is to provide a solution for this problem.